\documentclass{article}

\usepackage{arxiv}

\usepackage[utf8]{inputenc} % allow utf-8 input
\usepackage[T1]{fontenc}    % use 8-bit T1 fonts
\usepackage{hyperref}       % hyperlinks
\usepackage{url}            % simple URL typesetting
\usepackage{booktabs}       % professional-quality tables
\usepackage{amsfonts}       % blackboard math symbols
\usepackage{nicefrac}       % compact symbols for 1/2, etc.
\usepackage{microtype}      % microtypography
\usepackage{lipsum}		% Can be removed after putting your text content
\usepackage{graphicx}
\usepackage[super,sort&compress]{natbib}
\usepackage{doi}
\usepackage{amsmath,amssymb,mathtools}
\usepackage{caption}

\DeclareMathOperator{\logit}{logit}
\newcommand{\Se}{\mathrm{Se}}
\newcommand{\Sp}{\mathrm{Sp}}
\newcommand{\Ind}{\mathbb{I}}

\title{Methods for Assessing Distributional Assumptions in Individual Participant Data Meta-Analysis of Diagnostic Tests}

\author{ 
	\href{https://orcid.org/0000-0002-0171-588X}{\includegraphics[scale=0.06]{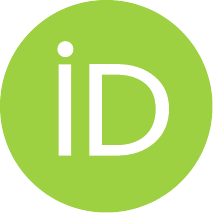}\hspace{1mm}Diribsa T. Bedada} \\
	Department of Statistics and Actuarial Science\\
	University of Waterloo\\
	Waterloo, Ontario, Canada \\
\And
\href{https://orcid.org/0000-0002-3154-3855}{\includegraphics[scale=0.06]{orcid.pdf}\hspace{1mm}Zelalem F. Negeri}\thanks{Corresponding author: Zelalem F. Negeri, Ph.D., Department of Statistics and Actuarial Science; University of Waterloo, M3 4116; Waterloo, Ontario, Canada N2L 3G1; Tel +1 519 888 4567, ext 4118; Email: znegeri@uwaterloo.ca.} \\
	Department of Statistics and Actuarial Science\\
	University of Waterloo\\
	Waterloo, Ontario, Canada
}

\renewcommand{\shorttitle}{Methods for Assessing Distributional Assumptions in IPDMA of DTAs}

\begin{document}
\maketitle

\begin{abstract}
Meta-analysis of diagnostic test accuracy (DTA) studies is commonly based on bivariate generalized linear mixed models (GLMMs) that assume normally distributed study-specific random effects. In practice, this distributional assumption is rarely examined, particularly in individual participant data meta-analysis (IPDMA), where empirical Bayes (EB) estimates exhibit heterogeneous precision across studies. Applying standard multivariate normality tests to such estimates can therefore lead to misleading inference. We propose a weighted extension of the Henze--Zirkler (HZ) test to assess bivariate normality of random effects, accounting for study-specific precision via inverse-variance weighting. To obtain valid finite-sample inference, we develop a parametric bootstrap calibration procedure. Extensive simulation studies showed that the classical HZ test applied to EB estimates can exhibit substantial Type I error inflation, whereas the proposed weighted procedures maintain nominal error rates while retaining power to detect departures from normality. The proposed method provides a practical, statistically principled diagnostic tool for assessing random-effects distributional assumptions in hierarchical models, with particular relevance to IPDMA of DTA studies. Although motivated by IPDMA, the proposed approach also extends to aggregate-data diagnostic test accuracy meta-analysis based on study-level $2\times2$ tables.
\end{abstract}

% keywords can be removed

\section{Introduction}
Generalized linear mixed models (GLMMs) are widely used for analyzing clustered and hierarchical data, with random effects introduced to capture between-cluster heterogeneity and induce correlation among observations \cite{breslow1993approximate, mcculloch2011prediction}. In multivariate settings, random-effects models allow correlated outcomes to be modelled jointly, improving efficiency and interpretability \cite{riley2007evaluation, jackson2011multivariate, riley2015multivariate}. A key assumption underlying these models is that random effects follow a multivariate normal distribution, which facilitates likelihood-based inference and prediction.

A prominent application arises in meta-analysis of diagnostic test accuracy
(DTA) studies, where joint modelling of sensitivity and specificity is required.
The bivariate random-effects model \cite{reitsma2005bivariate}, formulated
within a GLMM framework \cite{chu2006bivariate}, has become the standard
approach for both aggregate-data and individual participant data meta-analysis
(IPDMA). In particular, IPDMA offers increased flexibility in modelling
covariates and handling sparse data, but also introduces greater heterogeneity
in study-specific information \cite{debray2017guide, riley2010meta}.

The non-diagonal covariance structure allows study-specific random effects
for sensitivity and specificity to vary jointly rather than independently.
Such dependence may arise from between-study variation in
diagnostic thresholds: lower thresholds tend to increase sensitivity while
decreasing specificity, with the reverse occurring for higher thresholds.
The bivariate random-effects approach has its methodological roots in the
general bivariate meta-analysis framework of
\citet{van1993bivariate} and was subsequently formulated for DTA
meta-analysis by \citet{reitsma2005bivariate}. In DTA meta-analysis, the
bivariate model is closely related to the HSROC model of
\citet{rutter2001hierarchical}; \citet{harbord2007unification} established
the relationship between these two parameterizations. However, the
correlation in the bivariate model need not reflect threshold variation
exclusively, since other sources of between-study heterogeneity may also
induce dependence between sensitivity and specificity. Direct modelling of
SROC curves provides another class of approaches
\cite{holling2012meta,charoensawat2014meta}.

Despite its central role, the normality assumption for random effects is rarely assessed in practice. Departures from normality may arise from population heterogeneity, varying diagnostic thresholds, or latent subgroups, and can lead to biased estimates of heterogeneity and misleading predictive inference \cite{hamza2008binomial, jackson2012quantifying}.

In practice, random effects are typically estimated using empirical Bayes (EB) methods. These estimates are subject to shrinkage toward the population mean, with the degree of shrinkage depending on study-specific information such as sample size and within-study variability \cite{verbeke2000linear, mcculloch2011prediction}. Consequently, EB estimates exhibit heterogeneous precision and dependence, and therefore do not constitute independent and identically distributed realizations from the underlying random-effects distribution. This fundamentally complicates the use of standard diagnostic tools for assessing distributional assumptions in hierarchical models.

A common approach is to apply multivariate normality tests to the estimated random effects. Among available procedures, the Henze--Zirkler (HZ) test is attractive due to its affine invariance and good power properties \cite{henze1990class, henze1997extreme}. However, like other classical tests, it assumes independent and identically distributed observations, an assumption that is violated for EB estimates. As a result, applying unweighted tests may lead to misleading inference, particularly when study sizes and information vary substantially, as in IPDMA of DTA studies.

Motivated by this limitation, we propose a weighted extension of the HZ test that accounts for the heterogeneous precision of EB random-effects estimates. The proposed method incorporates inverse-variance weighting derived from within- and between-study variability, thereby aligning the test with the information structure of the underlying GLMM. Because the null distribution is not available in closed form, we use bootstrap procedures for calibration. Simulation studies and real-data applications demonstrate that the proposed method provides well-calibrated inference and offers a practical tool for assessing the normality assumption in hierarchical models, with particular relevance to DTA meta-analysis.

The remainder of the paper is organized as follows. Section~2 presents the motivating examples based on published meta-analyses. Section~3 introduces the bivariate GLMM for IPDMA of DTA studies and describes the classical and weighted HZ tests, as well as the proposed bootstrap calibration procedure. Sections~4 and~5 report the results of the simulation study and the real-data applications, respectively. Section~6 concludes with a discussion of practical implications and potential extensions.

\section{Motivating Examples}

In this section, we introduce two datasets to motivate the proposed methodology: one published meta-analytic aggregate dataset and one synthetic IPD dataset generated to mimic the characteristics of another published meta-analytic aggregate dataset. These examples are chosen to illustrate how the conclusions of normality assessment may vary depending on the characteristics of the data and the method used, highlighting the practical implications of accounting for heterogeneous precision and estimation uncertainty in EB estimates.

For each dataset, we examine the bivariate normality of the EB estimates of the random effects on the logit scale of sensitivity and specificity using the classical HZ test \cite{henze1990class}. The test was implemented using the \texttt{HZ.test} function from the \texttt{mvnTest} package in \textsf{R} \cite{pya2016mvntest}. In addition, we present univariate density plots with overlaid normal density curves for the EB estimates of the random effects on the logit scale of sensitivity and specificity to provide preliminary insight into potential departures from normality and to motivate the need for joint (bivariate) assessment.

\subsection{Cochrane hepatocellular carcinoma data}

The first example includes 147 studies evaluating the diagnostic
accuracy of abdominal ultrasound and alpha-fetoprotein for detecting
hepatocellular carcinoma in adults with chronic liver disease,
as reported in a Cochrane systematic review
\cite{colli2021abdominal}. Study-level $2\times2$ contingency
tables containing true-positive, false-positive, false-negative,
and true-negative counts were used for the analysis.
The dataset is provided in Ancillary File A (Table A.1). On average, each study includes approximately 355 participants, comprising about 139 diseased and 215 non-diseased individuals, indicating variability in both study size and the distribution of diseased and non-diseased groups across studies.

The fitted BGLMM yields a pooled sensitivity of approximately \(0.59\) and a pooled specificity of approximately \(0.84\) on the probability scale. The estimated between-study heterogeneity is small for sensitivity (variance \(= 0.17\)) and moderate for specificity (variance \(= 0.73\)), indicating greater variability across studies in specificity than in sensitivity. The estimated covariance between the random effects is negative (\(-0.27\)), suggesting a weak inverse association between sensitivity and specificity across studies.

\begin{figure*}[!t]
\centering

% Left: HCC dataset
\begin{minipage}[t]{0.49\textwidth}
    \centering
    \includegraphics[width=\linewidth]
    {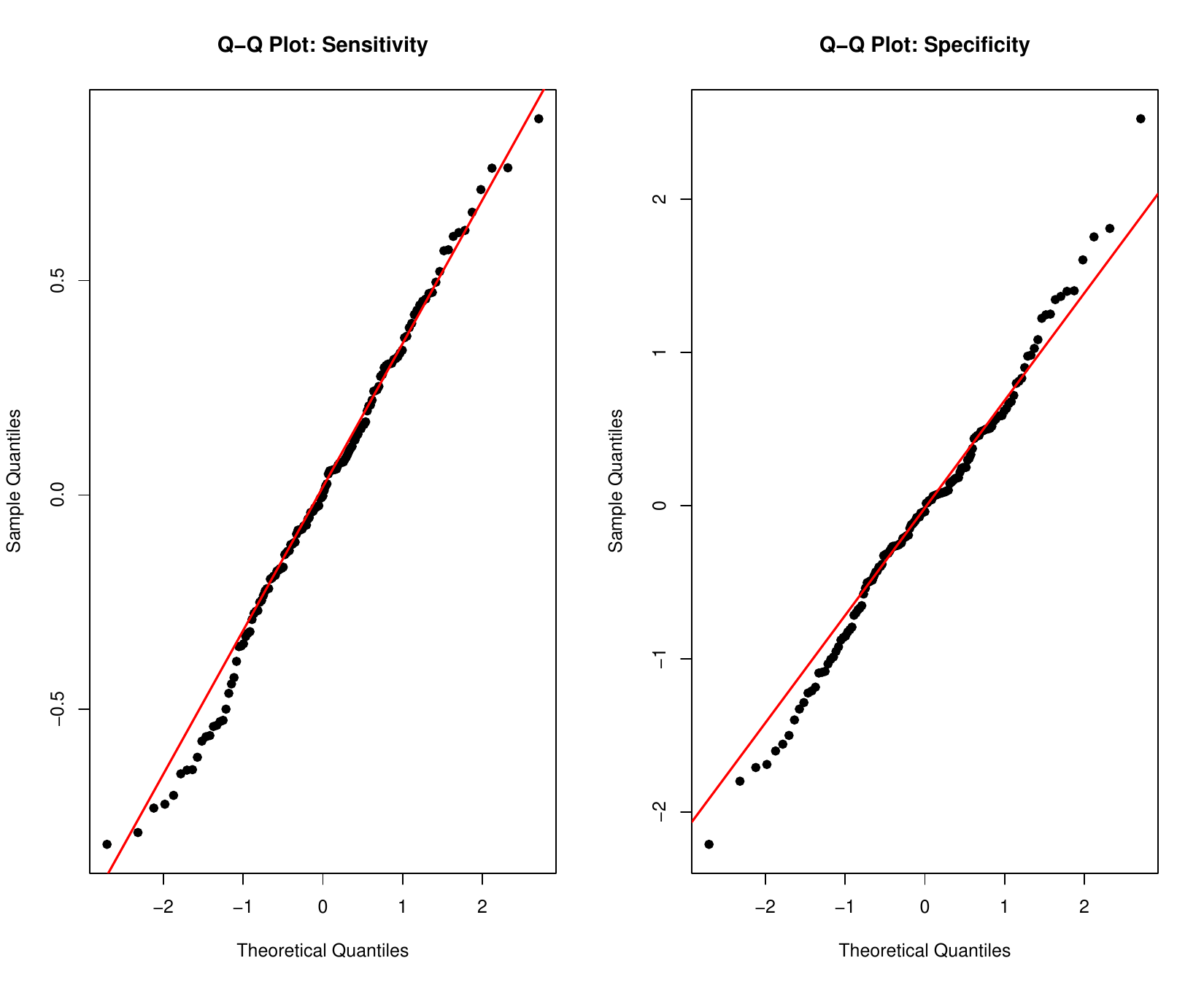}
    \captionof{figure}{Normal Q-Q plots of the EB estimates of the random effects for logit-transformed sensitivity and specificity, based on the hepatocellular carcinoma dataset extracted from the Cochrane review by \citet{colli2021abdominal}.}
    \label{fig:cochrane_QQ}
\end{minipage}
\hfill
% Right: SCOFF dataset
\begin{minipage}[t]{0.49\textwidth}
    \centering
    \includegraphics[width=\linewidth]
    {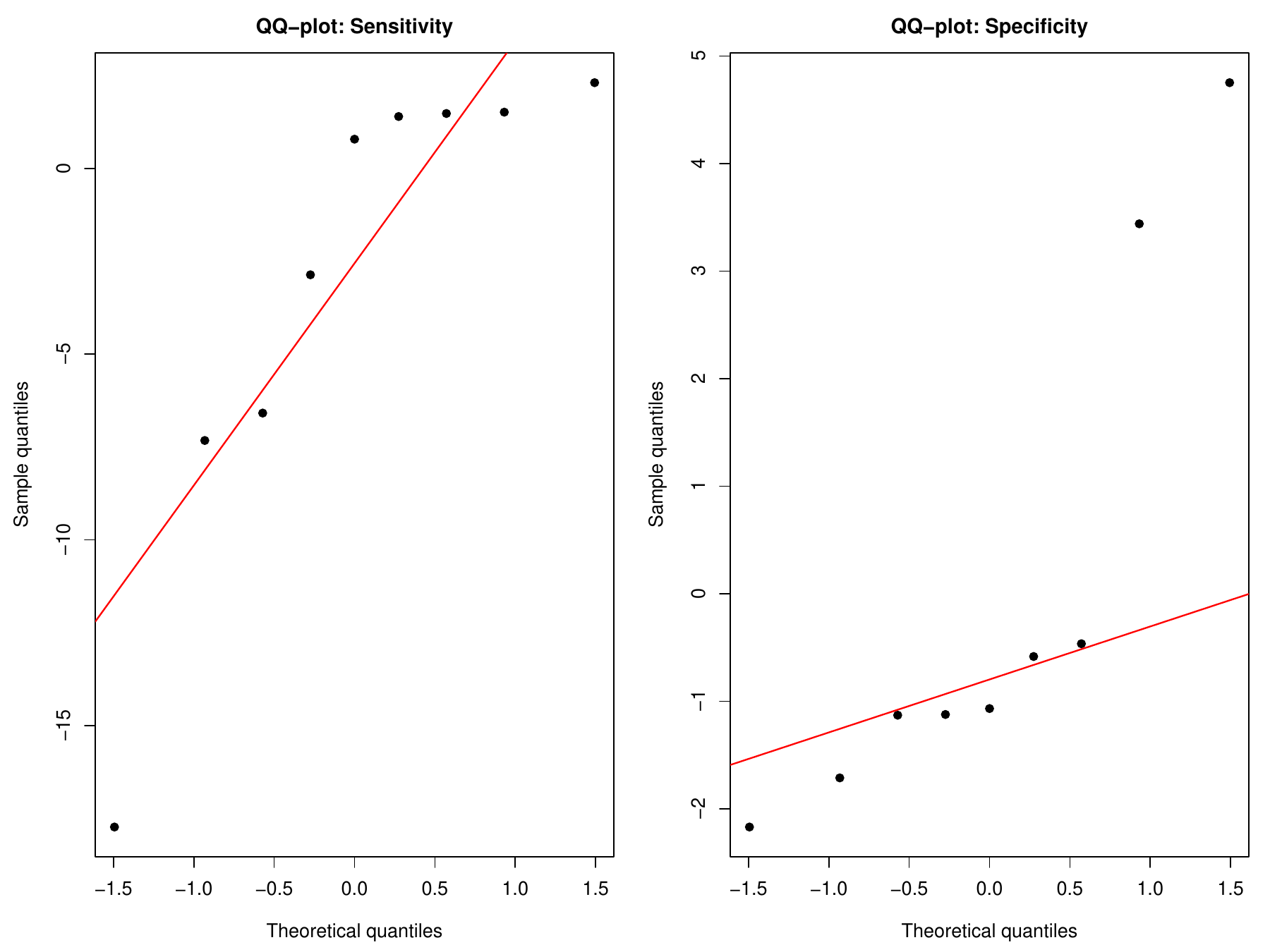}
    \captionof{figure}{Normal Q--Q plots of the EB
estimates of the study-specific random effects for
logit-transformed sensitivity and specificity, obtained
from the covariate-adjusted BGLMM fitted to the synthetic
SCOFF individual participant dataset generated using the
study structure reported by \citet{botella2015meta}.}
    \label{fig:scoff_QQ}
\end{minipage}

\end{figure*}

The empirical distributions of the EB estimates for logit-transformed sensitivity and specificity are shown in Figure~\ref{fig:cochrane_QQ}. Visual inspection of the Q-Q plots and normal density plots (see Figure B.1, Appendix B) indicates that the random effects for both sensitivity and specificity are approximately symmetric and closely follow the reference line, with only minor deviations in the tails. These patterns suggest that the normality assumption provides a reasonable approximation for the distribution of the random effects. To further investigate this, we applied the classical HZ test to assess multivariate normality. The test did not reject the null hypothesis of multivariate normality ($p$-value = 0.44), indicating no strong evidence against multivariate normality. 

\subsection{SCOFF questionnaire data}

The second example is based on a meta-analysis evaluating
the diagnostic accuracy of the SCOFF questionnaire for detecting
eating disorders \cite{botella2015meta}. Of the 15 studies,
9 employing a clearly defined reference standard were selected
to improve comparability. 
We generated synthetic IPD mimicking
the characteristics of these studies, including participant-level
age and sex covariates and generating random-effects from a mixture of two normal disributions to deviate from model assumptions. Details of the data-generating mechanism are provided in
Appendix A.
The synthetic dataset is provided in
Ancillary File A (Table A.2). On average, each study includes approximately 286 participants, comprising about 63 diseased and 223 non-diseased individuals, indicating a notable imbalance between the groups.

The covariate-adjusted BGLMM fitted to the synthetic
SCOFF IPD dataset yields a pooled sensitivity of
approximately \(0.932\) and a pooled specificity of
approximately \(0.967\), evaluated at the centered
age and sex covariates. The estimated between-study
heterogeneity is substantial for both sensitivity
(variance \(=15.240\)) and specificity
(variance \(=9.466\)), indicating considerable
variability in diagnostic accuracy across studies.
The estimated covariance between the study-specific
random effects is negative (\(-8.484\)), corresponding
to a correlation of approximately \(-0.706\).
This negative association is consistent with the
two-component mixture distribution used to generate
the synthetic random effects, whose opposing component
means induce negative dependence between logit
sensitivity and logit specificity.

The Q--Q plots (Figure~\ref{fig:scoff_QQ}) and density
plots (Figure B.2, Appendix B) reveal noticeable
departures from normality, particularly for specificity.
These graphical findings are consistent with both the
classical ($p=0.003$) and weighted ($p=0.032$) HZ tests,
which reject bivariate normality at the 5\% level. The results of the proposed test for these two datasets are presented in Section~5.

\section{Model and Proposed Method}

Consider $K$ studies indexed by $i=1,\ldots,K$. Within study $i$, let $y_{1im}\in\{0,1\}$ denote the indicator of a correct test result among diseased participants ($D_{im}=1$), for $m=1,\ldots,n_{1i}$; and let $y_{0ij}\in\{0,1\}$ denote the indicator of a correct test result among non-diseased participants ($D_{ij}=0$), for $j=1,\ldots,n_{0i}$. Thus, $y_{1im}=1$ corresponds to a true positive (TP) and $y_{1im}=0$ to a false negative (FN), while $y_{0ij}=1$ corresponds to a true negative (TN) and $y_{0ij}=0$ to a false positive (FP).

The conditional models are
\begin{equation}
\begin{aligned}
    &y_{1im} \mid u_{1i} \sim \mathrm{Bernoulli}(p_{1im}), \\
    &y_{0ij} \mid u_{0i} \sim \mathrm{Bernoulli}(p_{0ij}),
\end{aligned}
\label{eq:glmm_ipd}
\end{equation}
for $i=1,\ldots,K$, $m=1,\ldots,n_{1i}$, and
$j=1,\ldots,n_{0i}$.
In Equation~\eqref{eq:glmm_ipd}, $p_{1im}$ and $p_{0ij}$
denote the conditional probabilities of a positive test
result among diseased and a negative test result among
non-diseased participants, respectively.
These probabilities are modelled as
\begin{equation}
\begin{aligned}
\logit(p_{1im}) &= \beta_1 + u_{1i}
+ \gamma_{A,1}A_{1im}^\ast + \gamma_{S,1}S_{1im},\\
\logit(p_{0ij}) &= \beta_0 + u_{0i}
+ \gamma_{A,0}A_{0ij}^\ast + \gamma_{S,0}S_{0ij},
\end{aligned}
\label{eq:linpred_cov}
\end{equation}
where $\beta_1$ and $\beta_0$ in Equation~\eqref{eq:linpred_cov} are population-average intercepts for logit-sensitivity and logit-specificity, respectively, and $(\gamma_{A,1},\gamma_{S,1})$ and $(\gamma_{A,0},\gamma_{S,0})$ are covariate effects in diseased and non-diseased groups.

This participant-level formulation permits individual-specific covariates
to enter directly into the sensitivity and specificity components, a
feature not available from study-level $2\times2$ tables alone.

The study-specific random effects are
\[
\mathbf{u}_i=
\begin{pmatrix}
u_{1i}\\
u_{0i}
\end{pmatrix}
\sim
\mathcal{N}_2\!\left(
\begin{pmatrix}
0\\
0
\end{pmatrix},
\boldsymbol{\Sigma}
\right),
\qquad
\boldsymbol{\Sigma}=
\begin{pmatrix}
\sigma_1^2 & \sigma_{12}\\
\sigma_{12} & \sigma_2^2
\end{pmatrix}.
\]
Consequently,
\begin{align*}
\mathbb{E}\!\left[\logit(p_{1im}) \mid A_{1im}^\ast,S_{1im}\right]
&= \beta_1 + \gamma_{A,1}A_{1im}^\ast + \gamma_{S,1}S_{1im}, \\
\mathbb{E}\!\left[\logit(p_{0ij}) \mid A_{0ij}^\ast,S_{0ij}\right]
&= \beta_0 + \gamma_{A,0}A_{0ij}^\ast + \gamma_{S,0}S_{0ij}.
\end{align*}
and the between-study variability in $(\logit(\Se_i), \logit(\Sp_i))^\top$ is governed by $\boldsymbol{\Sigma}$.

The full likelihood and log-likelihood are provided in Appendix C. 
The log-likelihood in Equation 2 (Appendix C) does not have a closed-form expression due to the integration over the bivariate random effects, and likelihood-based inference therefore relies on numerical approximation. 
In this work, we employ adaptive Gauss--Hermite quadrature (AGHQ) to approximate the log-likelihood. 
Model estimation was carried out using the \texttt{mixed\_model} function from the \texttt{GLMMadaptive} R package \cite{rizopoulos2023glmmadaptive}.

\subsection{Henze--Zirkler tests for bivariate normality of random effects}
\subsubsection{Classical Henze--Zirkler test}
Let $\hat{\mathbf{u}}_1,\ldots,\hat{\mathbf{u}}_K\in\mathbb{R}^d$ denote estimated study-level random effects, that is, EB or best linear unbiased prediction (BLUP) estimates, from the fitted GLMM, with $d=2$ in our application. Let $\bar{\mathbf{u}}$ and $\mathbf{S}$ denote the sample mean and covariance matrix of $\{\hat{\mathbf{u}}_i\}$. Define the standardized residuals
\[
\mathbf{z}_i=\mathbf{S}^{-1/2}(\hat{\mathbf{u}}_i-\bar{\mathbf{u}}),\qquad i=1,\ldots,K.
\]
The classical HZ test is based on a weighted $L^2$ distance between the empirical characteristic function of the standardized residuals and that of the multivariate normal distribution. The corresponding test statistic is
\begin{align*}
T_{\mathrm{HZ}} = K \Bigg[
&\frac{1}{K^2}\sum_{i=1}^{K}\sum_{j=1}^{K}
\exp\!\left(-\frac{\beta^2}{2}\|\mathbf{z}_i-\mathbf{z}_j\|^2\right) \\
&-\frac{2}{K(1+\beta^2)^{d/2}}
\sum_{i=1}^{K}
\exp\!\left(-\frac{\beta^2}{2(1+\beta^2)}\|\mathbf{z}_i\|^2\right) \\
&+\frac{1}{(1+2\beta^2)^{d/2}}
\Bigg]
\end{align*}
The statistic $T_{\mathrm{HZ}}$ measures the discrepancy between the empirical distribution of the standardized residuals and the multivariate normal distribution. The tuning parameter $\beta$ is given by
\[
\beta
=
\frac{\left\{(2d+1)K/4\right\}^{1/(d+4)}}{\sqrt{2}}.
\]
From a theoretical perspective, under the null hypothesis of multivariate normality, the asymptotic distribution of $T_{\mathrm{HZ}}$ is nonstandard. Specifically, as $K \to \infty$, $T_{\mathrm{HZ}}$ converges in distribution to an infinite weighted sum of independent chi-square random variables,
\[
T_{\mathrm{HZ}} \xrightarrow{d} \sum_{k=1}^\infty \delta_k Z_k^2,
\]
where $Z_k \sim \mathcal{N}(0,1)$ are independent standard normal random variables and $\{\delta_k\}_{k\ge1}$ are nonnegative constants depending on the dimension $d$ and the tuning parameter $\beta$. These constants correspond to the eigenvalues of a certain integral operator defined through the weight function of the test statistic \cite{henze1990class}. Because the limiting distribution involves an infinite weighted sum of chi-square random variables with unknown eigenvalues, it is analytically intractable. Consequently, critical values are typically obtained via simulation-based or approximation methods \cite{henze1990class}.

In the present setting, this challenge is further complicated by the use of EB estimates, which are neither independent nor identically distributed. As a result, the classical calibration of the HZ test may be unreliable, thereby motivating the development of a bootstrap-based procedure that explicitly accounts for estimation uncertainty.

\subsubsection{Proposed weighted Henze--Zirkler test}
To incorporate heterogeneous estimation precision, we define study-specific weight matrices 
\[
\widehat{\boldsymbol{W}}_i = \widehat{\boldsymbol{V}}_i^{-1},
\]
where
\[
\widehat{\boldsymbol{V}}_i = \widehat{\boldsymbol{\Psi}}_i + \widehat{\boldsymbol{\Sigma}} .
\]
Here, $\widehat{\boldsymbol{\Psi}}_i$ represents an approximation to the within-study sampling variance of the logit-transformed sensitivity and specificity (obtained via the delta method), while $\widehat{\boldsymbol{\Sigma}}$ denotes the estimated between-study covariance matrix of the random effects. The matrix $\boldsymbol{V}_i$ therefore approximates the marginal covariance of the EB estimates, reflecting both within-study sampling variability and between-study heterogeneity. Using the $2\times2$ table for study $i$, with continuity correction as needed,

\[
\widehat{\boldsymbol{\Psi}}_i =
\begin{pmatrix}
\frac{1}{\mathrm{TP}_i}+\frac{1}{\mathrm{FN}_i} & 0 \\
0 & \frac{1}{\mathrm{FP}_i}+\frac{1}{\mathrm{TN}_i}
\end{pmatrix}
\]
The weighted mean and weighted covariance matrix of the estimated random effects are defined, respectively, as
\[
\bar{\boldsymbol{u}}_w
=
\left(\sum_{i=1}^K \widehat{\boldsymbol{W}}_i\right)^{-1}
\left(\sum_{i=1}^K \widehat{\boldsymbol{W}}_i\widehat{\boldsymbol{u}}_i\right),
\]

\[
\mathbf{S}_w
=
\left(\sum_{i=1}^K \widehat{\boldsymbol{W}}_i\right)^{-1/2}
\left[
\sum_{i=1}^K
\widehat{\boldsymbol{W}}_i^{1/2}
(\widehat{\boldsymbol{u}}_i-\bar{\boldsymbol{u}}_w)
(\widehat{\boldsymbol{u}}_i-\bar{\boldsymbol{u}}_w)^\top
\widehat{\boldsymbol{W}}_i^{1/2}
\right]
\left(\sum_{i=1}^K \widehat{\boldsymbol{W}}_i\right)^{-1/2}.
\]
This weighted formulation ensures that studies with more precise random-effects estimates contribute more to the construction of the test statistic, while studies with greater estimation uncertainty are downweighted accordingly. The weighting is governed by the study-specific precision matrices $\widehat{\boldsymbol{W}}_i = \widehat{\boldsymbol{V}}_i^{-1}$, which reflect both within-study variability and between-study heterogeneity. To ensure that the resulting covariance matrix is well-defined and suitable for subsequent standardization, a symmetric formulation is obtained by applying square-root transformations to the precision matrices, yielding a symmetric positive-definite matrix. This avoids the potential asymmetry and indefiniteness that may arise from direct matrix weighting and guarantees that the resulting Mahalanobis distances are valid. Consequently, the proposed weighted HZ test is both numerically stable and theoretically coherent.

Define standardized weighted residuals
\[
\boldsymbol{z}_{iw}=\boldsymbol{S}_w^{-1/2}(\widehat{\mathbf{u}}_i-\bar{\boldsymbol{u}}_w),\qquad i=1,\ldots,K,
\]
and compute the proposed weighted test statistic $T_{\mathrm{wHZ}}$:
\begin{align*}
T_{\mathrm{wHZ}} = K \Bigg[
&\frac{1}{K^2}\sum_{i=1}^{K}\sum_{j=1}^{K}
\exp\!\left(-\frac{\beta^2}{2}\|\mathbf{z}_{iw}-\mathbf{z}_{jw}\|^2\right)
\nonumber\\
&\;-\frac{2}{K(1+\beta^2)^{d/2}}
\sum_{i=1}^{K}
\exp\!\left(-\frac{\beta^2}{2(1+\beta^2)}\|\mathbf{z}_{iw}\|^2\right)
+\frac{1}{(1+2\beta^2)^{d/2}}
\Bigg].
\end{align*}
When all studies have identical precision (i.e., $\widehat{\boldsymbol{W}}_i = \widehat{\boldsymbol{W}}$ for all $i$), the proposed statistic ($T_{\mathrm{wHZ}}$) reduces to the classical HZ test ($T_{\mathrm{HZ}}$).

\subsection{Bootstrap calibration of the weighted test}
Because $T_{\mathrm{wHZ}}$ depends on estimated random effects and weight
matrices, its null distribution is not available in closed form. We therefore
use bootstrap calibration. Let $T_{\mathrm{wHZ}}^{\mathrm{obs}}$ denote the
observed statistic and let $T_{\mathrm{wHZ}}^{(b)}$ denote the statistic from
bootstrap replicate $b$.

Of the $B$ attempted bootstrap replications, let $B_{\mathrm{eff}}\leq B$
denote the number of valid replications. A replication is considered valid
when the GLMM is successfully fitted and the EB estimates, weight matrices,
and resulting $T_{\mathrm{wHZ}}^{(b)}$ are all available. The bootstrap
$p$-value is
\begin{equation}
\widehat p_{\mathrm{boot}}
=
\frac{
1+\displaystyle\sum_{b=1}^{B_{\mathrm{eff}}}
\Ind\!\left(
T_{\mathrm{wHZ}}^{(b)}
\geq T_{\mathrm{wHZ}}^{\mathrm{obs}}
\right)}
{B_{\mathrm{eff}}+1}.
\label{eq:bootstrap_pvalue}
\end{equation}
The $+1$ correction prevents zero $p$-values arising solely from finite
Monte Carlo sampling
\cite{davison1997bootstrap,phipson2016permutation}.

\subsubsection{Parametric bootstrap}

To approximate the null distribution of the proposed weighted HZ statistic,
we generate bootstrap datasets under the null hypothesis of bivariate normal
random effects using the fitted GLMM. Specifically, the fitted fixed effects and covariance matrix are treated as
the null-generating parameter estimates, and study-specific random effects
are generated from
$\mathcal{N}_2(\mathbf{0},\widehat{\boldsymbol{\Sigma}})$. Each bootstrap dataset is generated using the original study structure and
covariates, after which the same GLMM is refitted and the weighted HZ
statistic is recomputed
\cite{efron1993introduction,davison1997bootstrap,
chernick2014introduction}. The parametric bootstrap procedure is implemented as follows:
\begin{enumerate}

\item Fit the bivariate random-intercepts GLMM to the observed data to obtain
the estimated fixed-effects vector $\widehat{\boldsymbol{\beta}}$ and the
estimated between-study covariance matrix
$\widehat{\boldsymbol{\Sigma}}$.

\item Compute the observed weighted HZ statistic
$T_{\mathrm{wHZ}}^{\mathrm{obs}}$ from the observed dataset.

\item For each bootstrap replication $b=1,\ldots,B$:

\begin{enumerate}

\item For each study $i=1,\ldots,K$, draw a study-specific random-effects
vector
\[
\mathbf{u}_i^{(b)}
=
\begin{pmatrix}
u_{1i}^{(b)}\\
u_{0i}^{(b)}
\end{pmatrix}
\sim
\mathcal{N}_2(\mathbf{0},\widehat{\boldsymbol{\Sigma}}).
\]

\item Retain the original study structure, including the number of
individuals per study, disease indicators, and observed covariate values.

\item For each individual $m$ in study $i$, generate a bootstrap outcome
from the fitted GLMM according to
\[
Y_{im}^{(b)}
\sim
\mathrm{Bernoulli}\!\left(p_{im}^{(b)}\right),
\]
where
\[
\logit\!\left(p_{im}^{(b)}\right)
=
\boldsymbol{x}_{im}^{\top}\widehat{\boldsymbol{\beta}}
+
\boldsymbol{z}_{im}^{\top}\boldsymbol{u}_i^{(b)}.
\]

\item Fit the same GLMM to the bootstrap dataset, extract the EB estimates
$\{\widehat{\boldsymbol{u}}_i^{(b)}\}$, and compute the corresponding
bootstrap statistic $T_{\mathrm{wHZ}}^{(b)}$.

\end{enumerate}

\item Use the empirical distribution of the $B_{\mathrm{eff}}$ valid bootstrap
statistics to approximate the null distribution of the test statistic and
compute the bootstrap $p$-value using
Equation~\eqref{eq:bootstrap_pvalue}.

\end{enumerate}

\section{Simulation Study}
\subsection{Simulation design}
We conducted an extensive simulation study to evaluate the finite-sample performance of the classical and proposed weighted HZ tests under both bivariate normal and mixture-normal random-effects distributions. The simulation scenarios varied according to the number of studies $(K)$, average study sample size $(n)$, disease prevalence $(\pi)$, the magnitude of between-study heterogeneity, and the correlation between sensitivity and specificity. To guide the specification of realistic parameter settings, we analyzed 561 meta-analyses of DTA studies published in the Cochrane Database of Systematic Reviews from 2011 to 2022. The simulation parameters were then chosen to reflect the empirical ranges observed in this collection of meta-analyses. The operating characteristics of the competing procedures were assessed in terms of empirical Type I error rate and rejection probability under non-normal alternatives (power). Simulation settings and data-generating mechanisms are described in detail in the Ancillary File, Appendix D.

Both methods were applied to each simulated dataset. The classical HZ test was applied to EB estimates of the study-specific random effects obtained from a bivariate random-intercepts GLMM fitted with covariate effects for age and sex. The proposed weighted HZ test was then applied to the same EB estimates and calibrated using a parametric bootstrap procedure. Bootstrap calibration was performed as described in Section~3.2.1.

In the bootstrap procedure, calibration was based on \(200\) bootstrap
replications. 
This choice balanced Monte Carlo precision against the
substantial computational cost of bootstrap calibration
within each of the 1,000 simulation replications.
Sensitivity analyses were also conducted to assess the
robustness of the results to different values of $B$;
larger values may be used in empirical applications
when greater Monte Carlo precision is desired.
For each simulation scenario, the procedure was repeated over \(1{,}000\)
Monte Carlo replications. Performance was summarized by the empirical
rejection rates of the competing methods at the \(5\%\) and \(10\%\)
significance levels. Under the bivariate normal random-effects model, these
rejection rates were interpreted as empirical Type I error rates, whereas
under the mixture-normal random-effects model, they were interpreted as
power.

\subsection{Simulation Results}

\begin{figure*}[!t]
\centering
\includegraphics[width=0.8\textwidth]{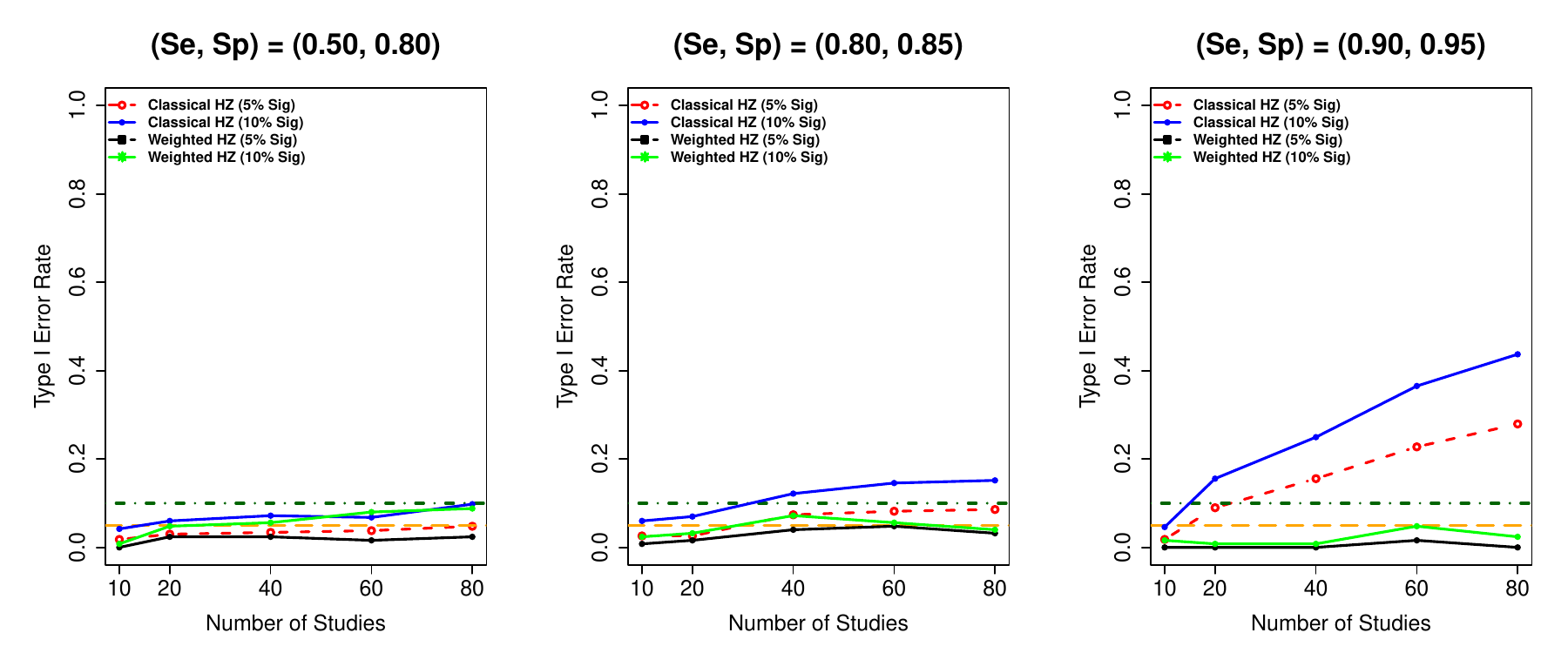}
\caption{Empirical Type I error rates of the classical and weighted Henze--Zirkler (HZ) tests under small sample sizes, with $(\sigma^{2}_{1}, \sigma^{2}_{2}, \sigma_{12}) = (0.40, 0.50, -0.02)$. Results are shown for nominal significance levels of 5\% and 10\%, with horizontal reference lines at 0.05 (orange) and 0.10 (dark green).}
\label{fig:type_I_error_p40_Figure1}
\end{figure*}

\begin{figure*}[!t]
  \centering
  \includegraphics[width=0.8\textwidth]{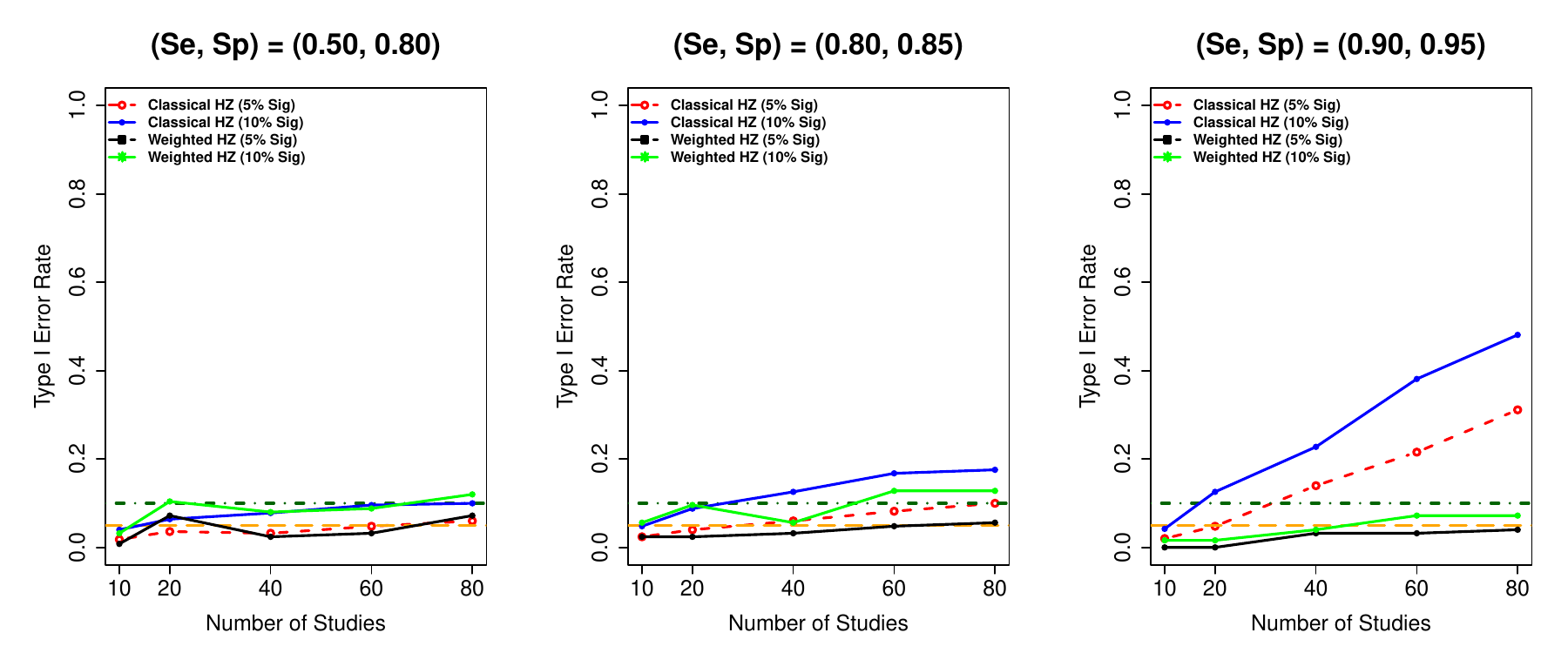}
  \caption{Empirical Type I error rates of the classical and weighted HZ tests when the number of patients in each group is small and $(\sigma^{2}_{1}, \sigma^{2}_{2}, \sigma_{12}) = (1.30, 2.02, -0.02)$. Results are shown for nominal significance levels of 5\% and 10\%, with horizontal reference lines at 0.05 (orange) and 0.10 (dark green).}
  \label{fig:type_I_error_p40_Figure5}
\end{figure*}

\begin{figure*}[!t]  
  \centering
  \includegraphics[width=0.8\textwidth]{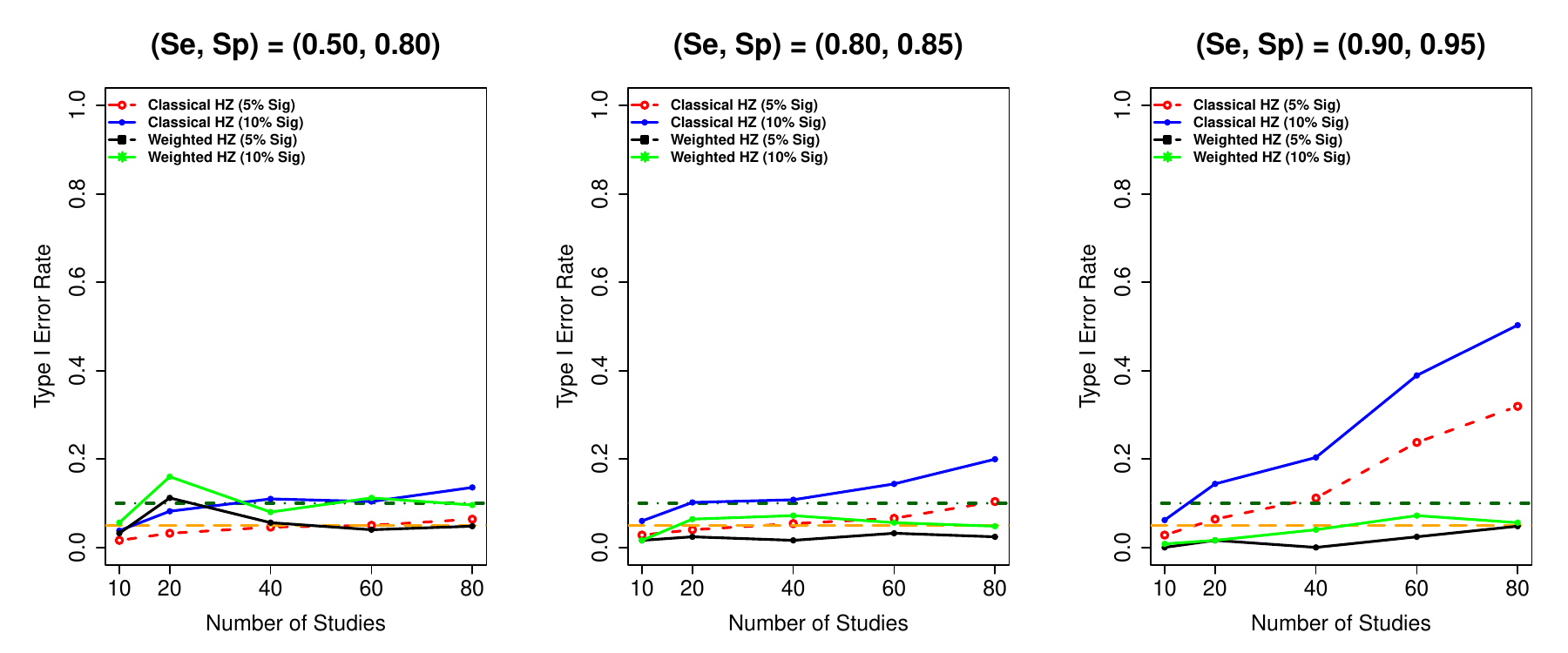}  
  \caption{Empirical Type I error rates of the classical and weighted HZ tests when the number of patients in each group is small and $(\sigma^{2}_{1}, \sigma^{2}_{2}, \sigma_{12}) = (1.30, 2.02, -0.18)$. Results are shown for nominal significance levels of 5\% and 10\%, with horizontal reference lines at 0.05 (orange) and 0.10 (dark green).}
 \label{fig:type_I_error_p40_Figure7}
\end{figure*}

\begin{figure*}[!t]  
  \centering
  \includegraphics[width=0.8\textwidth]{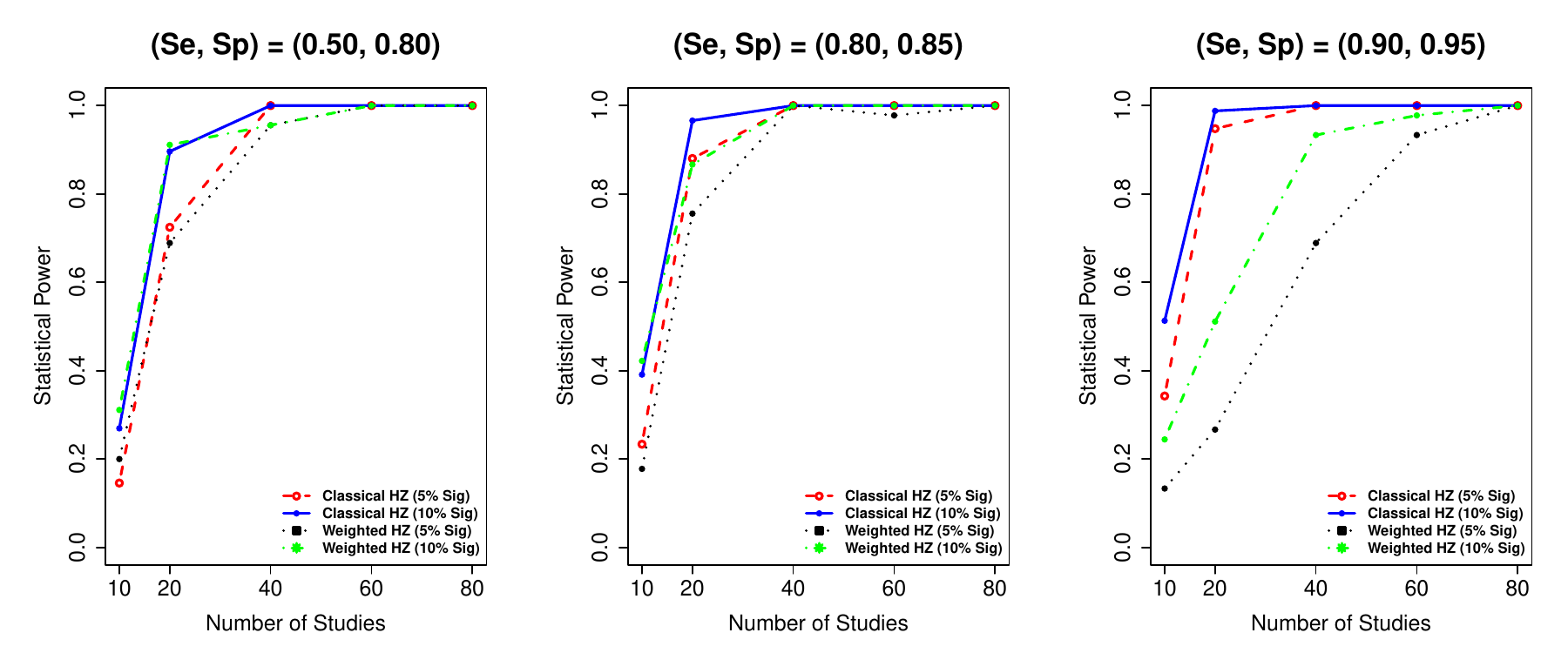}  
  \caption{Power of the classical and weighted HZ tests under the mixture alternative when the number of patients in each group is small and $(\sigma^{2}_{1}, \sigma^{2}_{2}, \sigma_{12}) = (0.40, 0.50, -0.02)$.}
 \label{fig:power_p40_Figure1}
\end{figure*}

\begin{figure*}[!t]  
  \centering
  \includegraphics[width=0.8\textwidth]{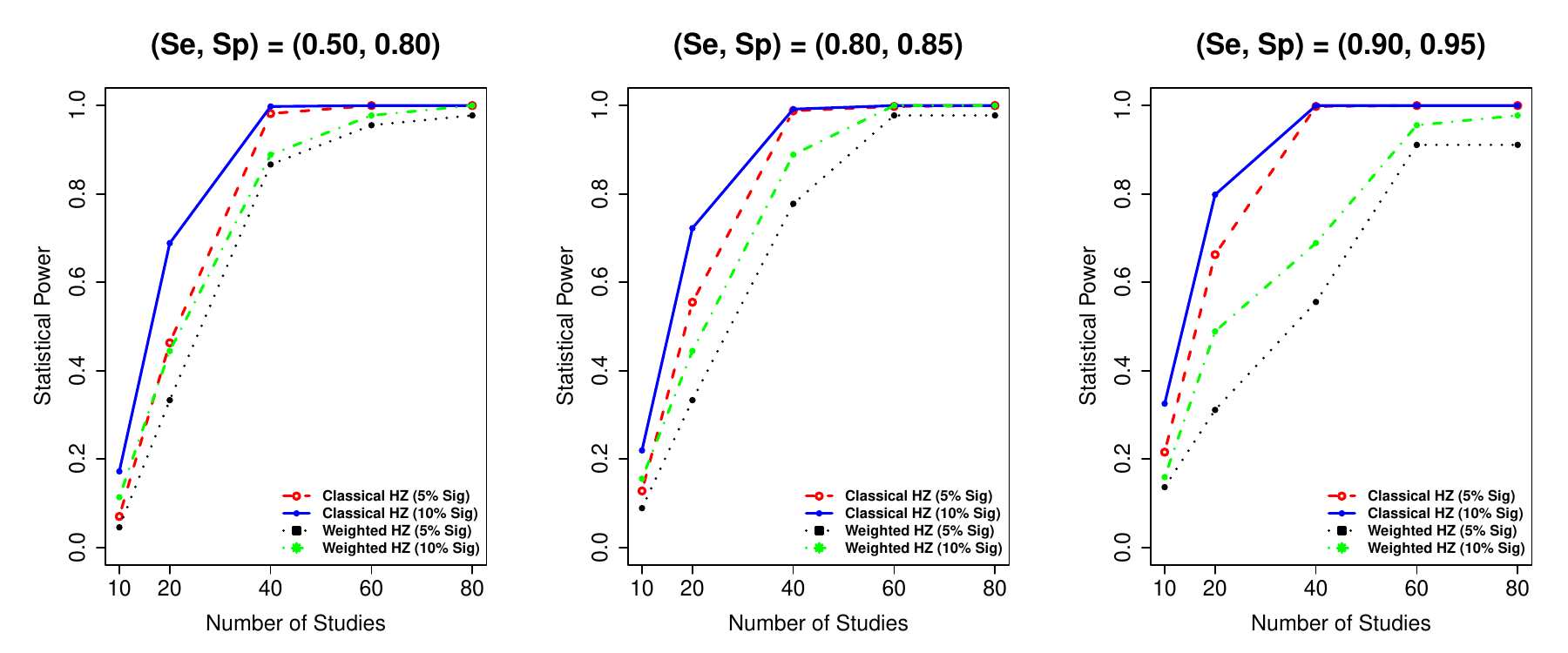}  
  \caption{Power of the classical and weighted HZ tests under the mixture alternative when the number of patients in each group is small and $(\sigma^{2}_{1}, \sigma^{2}_{2}, \sigma_{12}) = (1.30, 2.02, -0.02)$.}
   \label{fig:power_p40_Figure5}
 \end{figure*}

\begin{figure*}[!t]  
  \centering
  \includegraphics[width=0.8\textwidth]{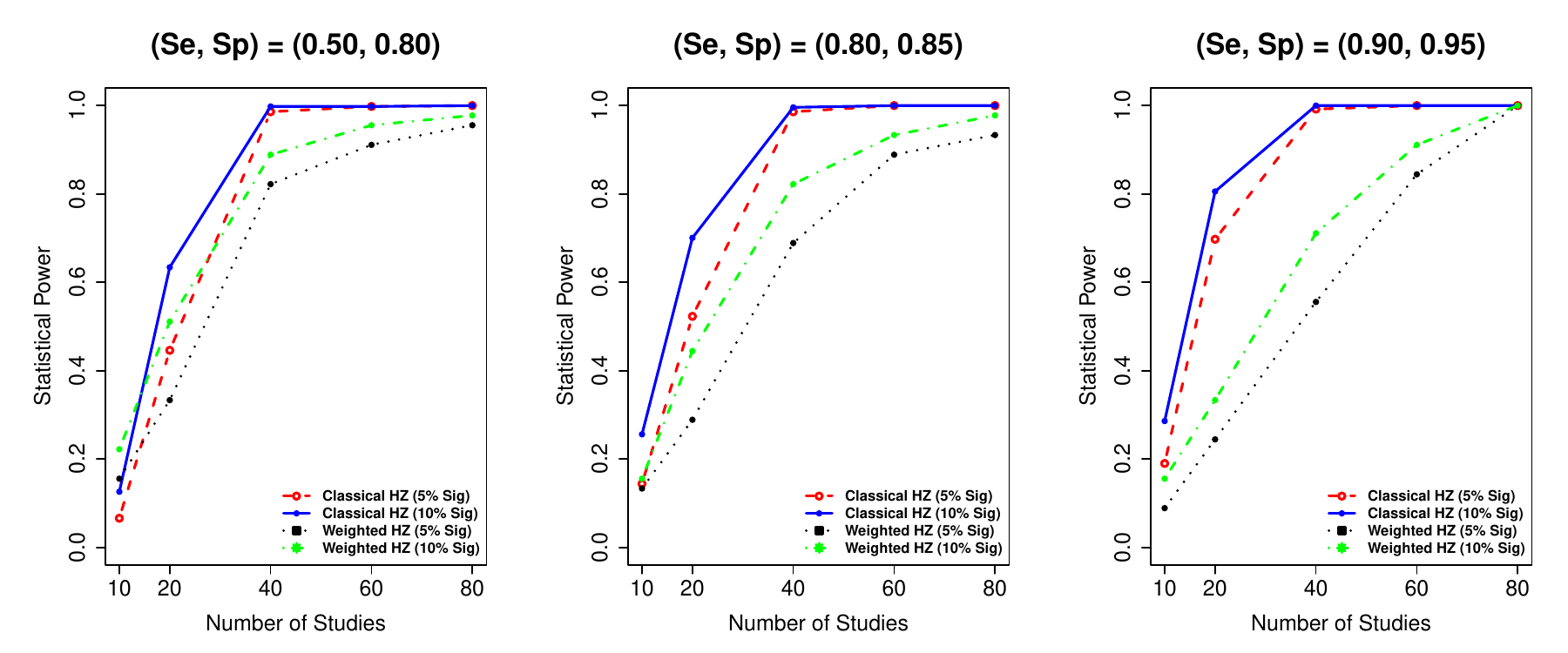}  
  \caption{Power of the classical and weighted HZ tests under the mixture alternative when the number of patients in each group is small and $(\sigma^{2}_{1}, \sigma^{2}_{2}, \sigma_{12}) = (1.30, 2.02, -0.18)$.}
 \label{fig:power_p40_Figure7}
\end{figure*}

We evaluated the finite-sample performance of the classical HZ test applied to EB estimates and compared it with the proposed weighted HZ test using a parametric bootstrap procedure across 240 simulation scenarios. The scenarios varied in the number of studies, the variances of the logit-transformed sensitivity and specificity, and the correlation between the two random effects. Performance was assessed in terms of empirical Type I error under a bivariate normal (BVN) random-effects model and statistical power under a mixture alternative designed to induce departures from normality. Simulation results for a prevalence of 20\% (Figures E.1--E.8, Appendix E, and Figures G.1--G.8, Appendix G) yielded the same overall conclusions as those for a prevalence of 40\%, presented in Figures~\ref{fig:type_I_error_p40_Figure1} --~\ref{fig:power_p40_Figure7} of the main manuscript, with additional results provided in Figures F.1--F.5 (Appendix F) and Figures H.1--H.5 (Appendix H). However, the magnitude of Type I error inflation and statistical power differed slightly across some scenarios.

Figure~\ref{fig:type_I_error_p40_Figure1} presents the empirical Type I error rates under scenarios with relatively small between-study variances, \((\sigma_1^2,\sigma_2^2)=(0.40,0.50)\), and weak covariance \((\sigma_{12}=-0.02)\). Under these conditions, the classical HZ test exhibits increasing inflation of the Type I error rate as the number of studies increases, particularly in settings with higher sensitivity and specificity. In contrast, the proposed weighted HZ test maintains rejection probabilities close to the nominal significance levels across all pairs of true sensitivity and specificity values, indicating good control of Type I error rates.

Figures~\ref{fig:type_I_error_p40_Figure5} and~\ref{fig:type_I_error_p40_Figure7} present the empirical Type I error rates under two covariance configurations of the bivariate normal random-effects model, differing only in the between-study covariance \(\sigma_{12}\), with prevalence fixed at $40\%$. In both scenarios, the between-study variances are large and the within-study sample sizes are small, allowing the effect of the dependence structure on test performance to be isolated.

In Figure~\ref{fig:type_I_error_p40_Figure5}, where the covariance is weak \((\sigma_{12}=-0.02)\), the classical HZ test already exhibits noticeable inflation of the Type I error rate, particularly as the number of studies increases. This inflation becomes substantially more pronounced in Figure~\ref{fig:type_I_error_p40_Figure7}, where the covariance is more negative \((\sigma_{12}=-0.18)\), corresponding to stronger negative dependence. Under this increased dependence, the Type I error rate of the classical HZ test rises sharply with the number of studies, reaching levels far exceeding the nominal significance thresholds. In contrast, the proposed weighted HZ test remains well calibrated in both settings, maintaining Type I error rates close to the nominal levels across all configurations.

Figure~\ref{fig:power_p40_Figure1} presents the statistical power under the mixture alternative when the between-study variances are relatively small, and the covariance is weak. As expected, the power of both procedures increases with the number of studies, reflecting improved ability to detect departures from normality as the sample size grows. The classical HZ test consistently exhibits higher power than the proposed test procedure, often approaching unity even for moderate numbers of studies. In contrast, the proposed approach shows a more gradual increase in power, particularly when the number of studies is small.

However, this apparent advantage of the classical HZ test should be interpreted with caution. As demonstrated in the Type I error analysis, the classical procedure does not maintain the nominal significance level and tends to over-reject under the null hypothesis. Consequently, the higher power observed here largely reflects this miscalibration rather than genuinely superior statistical power. In contrast, the weighted parametric bootstrap procedure provides a more reliable balance between Type I error control and power. This highlights the importance of evaluating power in conjunction with Type I error control, as procedures that fail to maintain nominal test levels may exhibit misleadingly high statistical power.

Figures~\ref{fig:power_p40_Figure5} and~\ref{fig:power_p40_Figure7} present the statistical power under the mixture alternative for scenarios with larger between-study variances. In both settings, both test procedures exhibit increasing power as the number of studies increases, reflecting improved ability to detect departures from normality with larger sample sizes.

The dependence structure between sensitivity and specificity also influences test performance. When the covariance is weak (Figure~\ref{fig:power_p40_Figure5}), the proposed weighted HZ attains moderate to high power, particularly as the variances increase. In contrast, when the covariance is more negative (Figure~\ref{fig:power_p40_Figure7}), power rates for the proposed approach are attenuated, indicating reduced sensitivity to departures from normality under stronger dependence. This behaviour can be explained by the geometry of the joint random-effects distribution. Strong negative dependence induces an elongated elliptical structure, with variability concentrated along a principal direction. After covariance-adjusted standardization, the distribution becomes more symmetric, making certain departures from normality less detectable and leading to lower power.

The magnitude of between-study variances also plays an important role. Larger variances reduce the impact of EB shrinkage, leading to less distortion in the estimated random effects. Although this improves the characteristics of both procedures, the classical HZ test remains less conservative, and its apparent power advantage persists even though it fails to control the Type I error rate.

One of the key findings from the simulation study is that, in several scenarios, the proposed weighted HZ test with parametric bootstrap demonstrates higher statistical power than the classical HZ test when the number of studies is small (see Appendices G  and H). This setting is particularly relevant in practice, as IPDMA of DTA studies often involves a limited number of studies. In such cases, the classical HZ test may exhibit reduced sensitivity to departures from normality because it relies on assumptions that are poorly aligned with the properties of EB estimates, including heterogeneity in precision and dependence across studies. In contrast, the proposed test statistic explicitly accounts for heterogeneous precision and estimation uncertainty, thereby improving the detection of distributional misspecification in small-sample settings. These findings highlight the practical advantage of the proposed method in realistic applications where the number of studies is limited.

\section{Real-life data application}

To formally assess multivariate normality in the motivating dataset introduced in Section~2, we applied both the classical HZ and the weighted HZ tests to the EB estimates.
For the HCC application, published study-level $2\times2$
tables were expanded into participant-level binary records
without individual covariates. For SCOFF, we used synthetic
IPD mimicking the original studies, including age and sex
covariates in the fitted BGLMM. To complement the distributional diagnostics,
Figures~B.3 and B.4 in Appendix~B of the Ancillary File present the BGLMM-based SROC curves for the
hepatocellular carcinoma and synthetic SCOFF datasets,
respectively, following \citet{chu2010bivariate} and
\citet{hamza2009multivariate}, with graphical implementation
adapted from \citet{dabi2026diagnostic}. The curves and
associated uncertainty regions illustrate differences in
sensitivity--specificity dependence and between-study
heterogeneity \cite{negeri2026structural}. Detailed
interpretations are provided in Appendix~B.

For the Cochrane hepatocellular carcinoma dataset, both the classical and weighted HZ tests failed to reject the null hypothesis of multivariate normality. 
This result is consistent with the graphical diagnostics and indicates that
the available data provide no evidence against the assumed bivariate normal
random-effects distribution. The corresponding results are reported in Table~\ref{tab:cochrane_results}.

\begin{table}[htbp]
\centering
\caption{Results of multivariate normality tests applied to EB estimates for the Cochrane hepatocellular carcinoma dataset.}
\begin{tabular}{l c c}
\hline
\textbf{Method} & \textbf{Statistic} & \textbf{$p$-value} \\
\hline
Classical HZ & 0.610 & 0.440 \\
Weighted HZ & 0.640 & 0.750 \\
\hline
\end{tabular}
\label{tab:cochrane_results}
\end{table}

For the synthetic SCOFF questionnaire IPD dataset,
both the classical and weighted HZ tests rejected
the null hypothesis of bivariate normality at the
5\% significance level. The classical HZ test
produced a statistic of 0.987 ($p=0.003$), whereas
the weighted HZ test yielded a statistic of 1.005
($p=0.032$). These findings provide evidence
against the bivariate normality assumption for
the study-specific random effects, consistent
with the two-component normal mixture distribution
used to generate the synthetic dataset.
The corresponding results are reported in
Table~\ref{tab:scoff_results}.

\begin{table}[htbp]
\centering
\caption{Results of multivariate normality tests
applied to empirical Bayes (EB) estimates from
the synthetic SCOFF questionnaire IPD dataset.}
\begin{tabular}{l c c}
\hline
\textbf{Method} & \textbf{Statistic} & \textbf{$p$-value} \\
\hline
Classical HZ & 0.987 & 0.003 \\
Weighted HZ & 1.005 & 0.032 \\
\hline
\end{tabular}
\label{tab:scoff_results}
\end{table}

Overall, the two applications illustrate contrasting
conclusions regarding bivariate normality. While
neither test rejected normality for the HCC dataset,
both detected departures in the synthetic SCOFF IPD
dataset, consistent with its underlying mixture
distribution.
 
\section{Discussion}

Assessing distributional assumptions for random effects is a fundamental yet often overlooked component of model diagnostics in hierarchical models. In GLMMs, the assumption of multivariate normal random effects underpins likelihood-based inference and summary measures \cite{chu2006bivariate}, but is rarely evaluated in practice, particularly in IPDMA of DTA studies where inference is based on EB estimates \cite{riley2008meta, breslow1993approximate}.

In this work, we examined the performance of multivariate normality tests applied to EB estimates and showed that the HZ test can be severely miscalibrated in this setting. Across simulation scenarios, the classical test exhibited substantial inflation of the Type I error rate, particularly under large between-study heterogeneity and extreme outcome probabilities. This behaviour arises because EB estimates are shrunken toward the population mean, exhibit heterogeneous precision, and are statistically dependent due to shared parameter estimation, and therefore do not constitute independent and identically distributed realizations from the underlying random-effects distribution \cite{verbeke2000linear, mcculloch2011prediction}.

To address this limitation, we proposed a weighted extension of the HZ test that accounts for heterogeneous estimation precision via inverse-variance weighting and uses bootstrap calibration. The proposed procedure aligns the test statistic with the hierarchical structure of the data and the statistical properties of EB predictors, and a simulation study showed that the proposed test consistently maintains control of the Type I error rate across all considered scenarios.

When departures from normality were introduced, the classical HZ test appeared to exhibit higher power. However, this reflects the same miscalibration that inflates Type I error under the null model. After proper calibration, the proposed weighted procedure achieves a more appropriate balance between test level and power, with power increasing as the number of studies increases. These findings highlight the distinction between apparent power and inferential validity.

The simulation results also illustrate how data characteristics influence performance. Strong dependence between sensitivity and specificity reduces the detectability of departures from normality, while smaller between-study variances increase the impact of EB shrinkage and further distort the distribution of the estimated random effects. These features reinforce the need for diagnostic procedures that account for heterogeneity, shrinkage, and dependence in hierarchical models, as also noted in multivariate meta-analysis \cite{jackson2011multivariate}.

From a practical perspective, the proposed method is simple to implement
after fitting standard BGLMMs and does not require modification of the
estimation framework.
A non-significant result from the proposed test should be interpreted as
insufficient evidence against the assumed random-effects distribution,
rather than as confirmation of normality.
Although motivated by IPDMA of DTA studies, the approach is broadly
applicable to other hierarchical and meta-analytic settings.

In summary, classical multivariate normality tests applied to EB estimates can lead to misleading conclusions in hierarchical models. The proposed weighted bootstrap HZ procedure provides a principled and practical alternative that yields reliable inference while accounting for heterogeneous precision, shrinkage, and dependence.

The proposed method has several limitations. It relies on model-based bootstrap, which may be sensitive to model misspecification, and operates on EB estimates rather than the unobserved random effects. The choice of weighting scheme may also influence performance, and the computational cost of bootstrapping may be non-negligible in large-scale settings.

Rejection of the bivariate normality assumption should prompt further
investigation rather than routine use of the normal random-effects model.
In aggregate data DTA meta-analysis, alternatives include flexible
parametric random-effects distributions
\cite{negeri2020skew,noma2022meta,negeri2020robust,
negeri2025bivariate} and nonparametric SROC approaches such as the
Lehmann-family model of \citet{holling2012meta}.
For IPDMA, we are developing a bivariate skew-normal GLMM that
accommodates asymmetric study-specific random effects while retaining
participant-level covariates \cite{bedada2026skewnormal}.
The proposed test in this manuscript is diagnostic and does not identify a unique
alternative; model choice should be guided by the nature of the
departure and scientific context. Future methodological research could extend the assessment
of random-effects distributional assumptions to alternative
DTA meta-analytic frameworks, including approaches that
directly model SROC curves
\cite{holling2012meta,charoensawat2014meta}, where random
effects are incorporated.

More broadly, these findings highlight the importance of assessing distributional assumptions for random effects in hierarchical models and the need for reliable diagnostic tools to guide the choice between standard normal models and more flexible alternatives.

\paragraph{Acknowledgments.}
The authors are grateful for the computing facilities provided by the Faculty of Mathematics at the University of Waterloo.

\paragraph{Funding Statement.}
ZFN would like to acknowledge funding from the Natural Sciences and Engineering Research Council of Canada (NSERC), grant number RGPIN-2024-04652.

\bibliographystyle{unsrtnat}
\bibliography{References}  %%% Uncomment this line and comment out the ``thebibliography'' section below to use the external .bib file (using bibtex) .

%%% Uncomment this section and comment out the \bibliography{references} line above to use inline references.
% \begin{thebibliography}{1}

% 	\bibitem{kour2014real}
% 	George Kour and Raid Saabne.
% 	\newblock Real-time segmentation of on-line handwritten arabic script.
% 	\newblock In {\em Frontiers in Handwriting Recognition (ICFHR), 2014 14th
% 			International Conference on}, pages 417--422. IEEE, 2014.

% 	\bibitem{kour2014fast}
% 	George Kour and Raid Saabne.
% 	\newblock Fast classification of handwritten on-line arabic characters.
% 	\newblock In {\em Soft Computing and Pattern Recognition (SoCPaR), 2014 6th
% 			International Conference of}, pages 312--318. IEEE, 2014.

% 	\bibitem{hadash2018estimate}
% 	Guy Hadash, Einat Kermany, Boaz Carmeli, Ofer Lavi, George Kour, and Alon
% 	Jacovi.
% 	\newblock Estimate and replace: A novel approach to integrating deep neural
% 	networks with existing applications.
% 	\newblock {\em arXiv preprint arXiv:1804.09028}, 2018.

% \end{thebibliography}

\end{document}